\documentclass[twocolumn,showpacs,preprintnumbers,amsmath,amssymb,nofootinbib]{revtex4}

\usepackage{graphicx}
\usepackage{dcolumn}
\usepackage{bm}
\usepackage{amssymb}
\usepackage{amsmath}
\usepackage{amsthm}

\begin{document}


\title{Next-to-leading order correction to the factorization limit of the radiation spectrum}

\author{M.V.~Bondarenco}
 \email{bon@kipt.kharkov.ua}
 \affiliation{%
NSC Kharkov Institute of Physics and Technology, 1 Academic St.,
61108 Kharkov, Ukraine }
 \affiliation{%
V.N. Karazine Kharkov National University, 4 Svobody Sq., 61077
Kharkov, Ukraine }

\date{\today}

\begin{abstract}

A next-to-leading order correction to the high-energy factorization
limit of radiation spectrum from an ultrarelativistic electron
scattering in an external field is evaluated. Generally, it does not
express through scattering characteristics, and accounts for
smoothness of the crossover between the initial and final electron
asymptotes. A few examples of application of this formula are given,
including bremsstrahlung in amorphous matter and undulator
radiation.


\end{abstract}





\maketitle

\section{Introduction}

It is well known \cite{Bloch-Nordsieck,Jauch-Rohrlich-IR} that
differential cross-section of bremsstrahlung in the limit of
vanishing photon frequency splits into a product of the differential
cross-section of elastic scattering and the photon emission
probability. The latter is essentially determined by classical
electrodynamics: even though the underlying scattering process may
be totally quantal, the radiation in the limit $\omega\to0$ is
predominantly generated on large distances during electron
rectilinear motion before and after the scattering, and rectilinear
motion is always semiclassical.

For ultrarelativistic radiating particles, the mentioned infrared
factorization  theorem was later superseded by another one
\cite{BK-ZhETF1968,Cheng-Wu-fact-theorem,Bjorken-Kogut-Soper,Akh-Ber}
stating that essentially the same kind of factorization may hold
even at $\hbar\omega\sim E$, granted that its actual condition is
the smallness of the target extent $T$ compared to the photon
formation length\footnote{Focusing on the ultrarelativistic case, we
let $c=1$.}
\[
l_f(\omega)=\frac{2EE'}{m^2\omega}, \qquad E'=E-\hbar\omega.
\]
The ratio $\frac{T}{l_f(\omega)}=\frac{m^2}{2EE'}\omega T$ can be
small even at $\hbar\omega\sim E$, provided $E\gg\frac{m^2
T}{\hbar}$. The latter condition is usually well satisfied for
ultrarelativistic electrons and microscopic scattering objects (when
$T\sim r_{B}=\hbar^2/me^2=137\hbar/m$), but can break down for
macroscopic targets. It may be more appropriate, therefore, to speak
here formally about the limit $T/E\to0$ rather than $\omega\to0$.

Case $\hbar\omega\sim E$ may actually be still treated
semiclassically, in spite that the differential radiation
probability under substantial photon recoil is influenced by
electron spin flips. Spin effects can be incorporated by means of
quantum electrodynamics. More importantly, at $E\to\infty$ the
electron wavelength $\hbar/E$ becomes short enough for the
scattering process to be semiclassical. That definitely must be so
in macroscopic external fields, but may hold as well in microscopic
ones, provided the final electron is not detected, wherewith
interference between different impact parameters disappears
\cite{Schwinger,BK-ZhETF1968,BKS,BLP}. The transferred momentum then
becomes a definite function of the impact parameter, and for each
impact parameter there exists a definite radiation emission
probability, as in classical electrodynamics.

Away from the infrared limit, shapes of semiclassical radiation can
be very diverse, reflecting the diversity of possible electron
motions in external fields. To make contact with this complexity, it
may be valuable to determine corrections to the infrared limit, as
$\omega T$ departs from zero. A step toward this goal was made long
ago by Low \cite{Low}, who had shown that the next-to-leading order
(NLO) infrared correction to the bremsstrahlung amplitude expresses
through the energy derivative of the elastic scattering amplitude at
fixed momentum transfer, i.e., is still determined by scattering
characteristics. It is easy to see, however, that such a procedure
gives a correction of order $\hbar\omega/E$ (involving $\hbar$, and
being insensitive to the external field length and strength), i.e.,
precisely that covered by the modified factorization theorem
\cite{BK-ZhETF1968,Cheng-Wu-fact-theorem,Bjorken-Kogut-Soper}.
Furthermore, in the ultrarelativistic QED case, when the elastic
scattering amplitude depends on the collision energy just linearly,
its nontrivial part depending on the momentum exchange with the
target will factor out, anyway, reducing the content of the Low's
theorem to that of the modified factorization theorem. Thus, for
ultrarelativistic particles such an approach seems to add
essentially nothing new,\footnote{Historically, however, Low's paper
\cite{Low} preceded generalized factorization theorem
\cite{BK-ZhETF1968,Cheng-Wu-fact-theorem,Bjorken-Kogut-Soper}.} and
be rather kinematical than related to the electron dynamics within
the target.


In practice, besides that, it often appears that typical radiation
emission angles from high-energy electrons are far too small for
experimental resolution, so one is content to measurement of the
angle-integral radiation spectrum. The task of deriving for it a
NLO infrared correction may be not so straightforward, because in
the framework of classical particle description, $\omega$ dependence
enters to the exponent of a plane wave, whereas expanding this
exponential to power series and integrating termwise may lead to
divergent or improper integrals.

The aim of the present article is to demonstrate that the NLO
correction to the angle-integral radiation spectrum has the order
$\mathcal{O}\left[T/l_f(\omega)\right]$ as $T/l_f\to0$, and to
derive for it a formula valid for arbitrary external field and
$\hbar\omega\sim E$, presuming the electron to be ultrarelativistic.
It will then be instructive to discuss its physical meaning, sign
and magnitude for several physical processes.

\section{Double time integral representation}

As has already been mentioned, the infrared limit of bremsstrahlung
emission probability may be inferred from classical electrodynamics.
The generic representation for the spectral-angular distribution of
classical radiation reads \cite{LL,Jackson}
\begin{subequations}\label{spectral-angular}
\begin{eqnarray}
\frac{dI}{d\omega d^2n}
&=&e^2\left|\frac{\omega}{2\pi}\int_{-\infty}^{\infty}dt
e^{i\omega\left[t-\bm{n}\cdot \bm{r}(t)\right]}\bm{n}\times\bm{v}(t) \right|^2 \label{init-t-int}\\
&\equiv&e^2\left|\frac{1}{2\pi}\int_{-\infty}^{\infty}dt
e^{i\omega\left[t-\bm{n}\cdot \bm{r}(t)\right]}\frac{d}{dt}
\frac{\bm{n}\times\bm{v}(t)}{1-\bm{n}\cdot \bm{v}(t)} \right|^2.
\label{init-t-int-ddt}
\end{eqnarray}
\end{subequations}
At $\omega\to0$ the exponential factor within the domain of action of the external field may be neglected, whereupon the time integral is trivially taken to give
\begin{eqnarray}\label{BH-semicl}
\frac{dI_{\text{BH}}}{d\omega}&=&\frac{e^2}{(2\pi)^2} \int d^2n
\left|\frac{\bm{n}\times\bm{v}_f(t)}{1-\bm{n}\cdot \bm{v}_f(t)}
-\frac{\bm{n}\times\bm{v}_i(t)}{1-\bm{n}\cdot \bm{v}_i(t)}
\right|^2 \nonumber\\
&=&\frac{2e^2}{\pi}\left(\frac{2+\gamma^2v_{fi}^2}{\gamma v_{fi}\sqrt{1+\gamma^2 v_{fi}^2/4}}\text{arsinh}\frac{\gamma v_{fi}}{2}-1\right).
\end{eqnarray}
It is a function of single variable $\gamma v_{fi}$, where $\bm{v}_{fi}=\bm{v}_f-\bm{v}_i$
[with $\bm{v}_i=\bm{v}(-\infty)$, $\bm{v}_f=\bm{v}(+\infty)$] is the electron scattering angle.

At $\hbar\omega\sim E$, quantum electrodynamics gives
\begin{eqnarray}\label{BH-quant}
\frac{dI_{\text{BH}}}{d\omega}=\frac{2e^2}{\pi}\left[\frac{2m\left(1+\frac{E^2+E'^2}{EE'}\frac{q^2}{4m^2}\right)}{q\sqrt{1+\frac{q^2}{4m^2}}}\text{arsinh}\frac{q}{2m}-1\right],
\end{eqnarray}
where $q\simeq q_{\perp}$ is the momentum transfer to the target,
which is predominantly transverse ($q_z\ll q_{\perp}$).

The situation becomes more intricate when one aims to derive a
next-to-leading order correction to (\ref{BH-semicl}). If one
attempts to expand the radiation amplitude into power series in
$\omega$ via expanding the exponential in (\ref{spectral-angular})
into Maclaurin series,
\[
e^{i\omega
t-i\bm{k}\cdot\bm{r}(t)}=1+i\omega[t-\bm{n}\cdot\bm{r}(t)]+\mathcal{O}(\omega^2T^2),
\]
the square of the corresponding amplitude would give a real
$\mathcal{O}(\omega^2T^2)$ correction, but the angular integral from
it will diverge, making such an approach for the radiation spectrum
rather ineffectual. Hence, it may be inappropriate to this end to
expand the entire phase factor. As will be shown below, in fact, the
expansion of $\frac{dI}{d\omega}$ beyond the IR factorization limit
begins with a term $\mathcal{O}(\omega T)$.


Better suited for expansion of the spectrum in powers of $\omega T$ (or $T/l_f$) is representation
\cite{BKF,BKS,Lindhard-semicl-rad,Akh-Shul-PhysRep,Blankenbecler-Drell}
\begin{eqnarray}\label{dIdomega-t1t2}
\frac{dI}{d\omega}=\omega\frac{e^2}{\pi}\int_{-\infty}^{\infty}dt_2 \int_{-\infty}^{t_2}\frac{dt_1}{t_2-t_1} \qquad\qquad\quad\nonumber\\
\times \Bigg\{ \left(\gamma^{-2}+\frac{E^2+E'^2}{4EE'} \left[\bm{v}(t_2)- \bm{v}(t_1)\right]^2\right)\nonumber\\
\times
\sin \frac{\omega E}{E'} \left[t_2-t_1-\left|\bm{r}(t_2)-\bm{r}(t_1)\right|\right]\nonumber\\
-\gamma^{-2} \sin\frac{\omega E}{E'} (1-v)(t_2-t_1)\Bigg\},
\end{eqnarray}
where the trajectory describes a nonradiating particle with the
initial conditions of the incoming electron (having energy $E$),
while the block $(1-v)(t_2-t_1)+v(t_2-t_1)-\left|\bm{r}(t_2)-\bm{r}(t_1)\right|$ entering the argument of the first sine may be
expressed through transverse velocity components as
\begin{eqnarray}\label{vtau-|r2-r1|}
v(t_2-t_1)-|\bm{r}(t_2)-\bm{r}(t_1)|\qquad\qquad\qquad\qquad\qquad\nonumber\\
\simeq
\frac{1}{2v(t_2-t_1)}\left\{v^2(t_2-t_1)^2-\left[\int_{t_1}^{t_2}dt \bm{v}(t)\right]^2\right\}\nonumber\\
\simeq
\frac{1}{2v}\left\{\int_{t_1}^{t_2}dt\bm{v}_{\perp}^2(t)-\frac1{t_2-t_1}\left[\int_{t_1}^{t_2}dt
\bm{v}_{\perp}(t)\right]^2\right\}.
\end{eqnarray}
The latter expression is rotation invariant, but in practice, it may
be advantageous to recast it a manifestly rotation invariant form
\begin{eqnarray}
v(t_2-t_1)-|\bm{r}(t_2)-\bm{r}(t_1)|\qquad\qquad\qquad\qquad\qquad\quad\nonumber\\
= \frac{1}{2v} \Bigg\{ \frac1{t_2-t_1} \int_{t_1}^{t_2}dt
\left[\bm{v}_f-\bm{v}(t)\right]
\cdot\int_{t_1}^{t_2}dt \left[\bm{v}(t)-\bm{v}_i\right]\nonumber\\
-\int_{t_1}^{t_2}dt
\left[\bm{v}_f-\bm{v}(t)\right]\cdot\left[\bm{v}(t)-\bm{v}_i\right]
\Bigg\}\quad \label{vtau-chi-chi-b}
\end{eqnarray}
with arbitrary $\bm{v}_i$, $\bm{v}_f$. If $\bm{v}_i$ is chosen to be
the initial, and $\bm{v}_f$ the final electron velocity, the
velocity differences appearing in Eq.~(\ref{vtau-chi-chi-b}) will
vanish correspondingly at $t\to-\infty$ and at $t\to+\infty$,
ensuring the convergence of the integrals at large times.

Finally, by virtue of fair straightness of the electron trajectory
at high energy, it can as well be reexpressed through the trajectory
of an electron with the same impact parameter but energy $E'$, or,
more symmetrically, through momentum exchange with the target
$\bm{q}(t)=\int^t dt \bm{F}(t)$, which is accumulated continuously
regardless of whether the photon was emitted or not. In terms of the
latter, $\bm{q}(t)=E\bm{v}_{\perp}(t)$, the radiation spectrum
expresses as
\begin{eqnarray}\label{}
\frac{dI}{d\omega}=\omega\frac{e^2}{\pi \gamma^2}\int_{-\infty}^{\infty}dt_2 \int_{-\infty}^{t_2}\frac{dt_1}{t_2-t_1} \qquad\qquad\qquad\qquad\qquad\quad\nonumber\\
\times \Bigg\{ \left(1+\frac{E^2+E'^2}{4EE' m^2} \left[\bm{q}(t_2)-
\bm{q}(t_1)\right]^2\right)
\sin \frac{t_2-t_1+\Delta s(t_2,t_1)}{l_f(\omega)}\nonumber\\
-\sin\frac{t_2-t_1}{l_f(\omega)}\Bigg\},\quad
\end{eqnarray}
with
\begin{eqnarray}\label{}
\Delta s(t_2,t_1)=\frac{1}{m^2} \Bigg\{ \frac1{\tau}
\int_{t_1}^{t_2}dt \left[\bm{q}_f-\bm{q}(t)\right]
\cdot\int_{t_1}^{t_2}dt \left[\bm{q}(t)-\bm{q}_i\right]\nonumber\\
-\int_{t_1}^{t_2}dt
\left[\bm{q}_f-\bm{q}(t)\right]\cdot\left[\bm{q}(t)-\bm{q}_i\right]
\Bigg\}.\quad
\end{eqnarray}

\section{Factorization limit and NLO calculation}

The Bethe-Heitler limit results from Eqs.~(\ref{dIdomega-t1t2}),
(\ref{vtau-chi-chi-b}) if the electron trajectory is replaced by its
counterpart corresponding to an instantaneous momentum transfer equal to $q_{fi}$ (say, at time $t=0$):
\begin{eqnarray}\label{BH-t1t2}
\frac{dI_{\text{BH}}}{d\omega}=\omega\frac{e^2}{\pi \gamma^{2}}\int_{0}^{\infty}dt_2 \int_{-\infty}^{0}\frac{dt_1}{t_2-t_1} \qquad\qquad\qquad\qquad\nonumber\\
\times \Bigg\{ \left(1+\frac{E^2+E'^2}{4EE'm^2}q_{fi}^2\right)\sin
\frac{\tau-\frac{t_1 t_2}{\tau}\frac{q_{fi}^2}{m^2}}{l_f(\omega)} -
\sin\frac{\tau}{l_f(\omega)} \Bigg\}.
\end{eqnarray}
Passing here to variables $\tau =t_2-t_1$, $w=t_2/\tau$
\cite{Bond-Shul-double-scat} reduces it to a single algebraic
integral
\begin{eqnarray*}\label{}
\frac{dI_{\text{BH}}}{d\omega}=\omega\frac{e^2}{\pi \gamma^{2}} \int_{0}^{1} dw \int_{0}^{\infty}d\tau \qquad\qquad\qquad\qquad\qquad\quad \nonumber\\
\times \Bigg\{ \left(1+\frac{E^2+E'^2}{4EE'm^2}q_{fi}^2\right)\sin  \frac{\left[1+\frac{q_{fi}^2}{m^2}w(1-w)\right]\tau}{l_f(\omega)} \nonumber\\
- \sin\frac{\tau}{l_f(\omega)} \Bigg\}\nonumber\\
=\frac{2e^2}{\pi}\frac{E'}{E}  \left\{ \int_{0}^{1} dw
\frac{1+\frac{E^2+E'^2}{4EE'm^2}q_{fi}^2}{1+\frac{q_{fi}^2}{m^2}
w(1-w)}-1\right\},\qquad\quad
\end{eqnarray*}
which is taken to give (\ref{BH-quant}). To evaluate a correction to
it, one has to subtract (\ref{BH-t1t2}) from (\ref{dIdomega-t1t2}),
and investigate the difference in the limit $ T/l_f(\omega)\to0$.

To this end, it is convenient to select a finite domain $0<t<T$,
containing all the deflecting fields, so that outside of it one may
let $\bm{q}(t\leq0)=\bm{q}_i$, $\bm{q}(t\geq T)=\bm{q}_f$. In the
double time integral, it is then possible to replace the lower limit
for $t_2$ by 0, and split the $t_1$ integral as
$\int_{-\infty}^{t_2}dt_1\ldots
=\int_{-\infty}^{0}dt_1\ldots+\int_{0}^{t_2}dt_1\ldots$. That gives
\[
\frac{dI}{d\omega}-\frac{dI_{\text{BH}}}{d\omega}=I_1+I_2
\]
with
\begin{eqnarray}\label{dIdomega-dIBHdomega}
I_1=\omega\frac{e^2}{\pi \gamma^{2}}\int_{0}^{\infty}dt_2 \int_{-\infty}^{0}\frac{dt_1}{\tau} \qquad\qquad\qquad\qquad\qquad \nonumber\\
\times \Bigg\{ \left(1+\frac{E^2+E'^2}{4EE'm^2} \left[\bm{q}(t_2)- \bm{q}_i \right]^2\right)\sin\frac{\tau+\Delta s(t_2,t_1)}{l_f(\omega)}\nonumber\\
-\left(1+\frac{E^2+E'^2}{4EE'm^2} q_{fi}^2\right)\sin
\frac{\tau-\frac{t_1t_2}{\tau m^2}q_{fi}^2}{l_f(\omega)} \Bigg\},
\end{eqnarray}
[the last term stemming from Eq.~(\ref{BH-t1t2})] and
\begin{eqnarray}
I_2=\omega\frac{e^2}{\pi \gamma^{2}}\int_{0}^{\infty}dt_2 \int_{0}^{t_2}\frac{dt_1}{\tau} \qquad\qquad\qquad\qquad\qquad\quad\nonumber\\
\times \Bigg\{\left(1+ \frac{E^2+E'^2}{4EE'm^2} \left[\bm{q}(t_2)- \bm{q}(t_1)\right]^2\right)\sin \frac{\tau+\Delta s(t_2,t_1)}{l_f(\omega)} \nonumber\\
- \sin\frac{\tau}{l_f(\omega)}\Bigg\}.\quad
\end{eqnarray}

In $I_1$, according to Eq.~(\ref{vtau-chi-chi-b}), the nonlinear part
of the phase of the first sine can be written
\begin{eqnarray}\label{t1t2+O}
\Delta s &=& \frac{1}{m^2} \Bigg\{ \frac1{\tau}
\left(-t_1\bm{q}_{fi}+\int_{-\infty}^{\infty}dt t\frac{d\bm{q}}{dt} \right)\nonumber\\
&\,&\qquad\qquad\qquad\cdot\left(t_2\bm{q}_{fi}-\int_{-\infty}^{\infty}dt t\frac{d\bm{q}}{dt} \right)\nonumber\\
&\,&\qquad\qquad - \int_{-\infty}^{\infty}dt \left[\bm{q}_f-\bm{q}(t)\right]\cdot\left[\bm{q}(t)-\bm{q}_i\right]\Bigg\}\nonumber\\
&=&-\frac{t_1t_2}{\tau
m^2}q_{fi}^2+\mathcal{O}\left(T{q_{fi}^2}/{m^2}\right),
\end{eqnarray}
which thus appears to be close to that in the second sine. Being
divided by $l_f(\omega)$, the first term in (\ref{t1t2+O}) can still
be nonvanishing as $T/l_f\to0$, because both typical contributing
times expand proportionally to $l_f(\omega)$, but
$\frac{T}{l_f}\frac{q_{fi}^2}{m^2}$ does vanish in this limit.
Therefore, there is complete cancellation between the last two lines
in Eq.~(\ref{dIdomega-dIBHdomega}) when $t_2>T$. In the difference
of those terms, concentrated at $t_2<T$, it is justified to neglect
$\Delta s/l_f$ in the phase of the first sine, wherewith this sine
factors out:
\begin{eqnarray}\label{}
I_1
\simeq \omega\frac{e^2}{\pi}\frac{E^2+E'^2}{4E^3E'}\int_{0}^{\infty}dt_2 \int_{-\infty}^{0}\frac{dt_1}{\tau} \qquad\qquad\nonumber\\
\times \left\{\left[\bm{q}(t_2)- \bm{q}_i \right]^2 -
q_{fi}^2\right\}\sin \frac{\tau}{l_f(\omega)}.
\end{eqnarray}
Now passage to variable $\varphi= {\tau}/{l_f(\omega)} $,
\begin{eqnarray}\label{}
I_1\simeq\omega\frac{e^2}{\pi}\frac{E^2+E'^2}{4E^3E'}\int_{0}^{\infty}dt_2
\left\{\left[\bm{q}(t_2)- \bm{q}_i \right]^2
- q_{fi}^2\right\} \nonumber\\
\times\int_{ {t_2}/{l_f(\omega)} }^{\infty}\frac{d\varphi}{\varphi} \sin
\varphi,\quad
\end{eqnarray}
proves that it tends to
\begin{eqnarray}\label{Iee-IBH}
I_1\underset{T/l_f(\omega)\to0}\to
\omega\frac{e^2}{2}\frac{E^2+E'^2}{4E^3E'}\qquad\qquad\qquad\qquad\nonumber\\
\times \int_{0}^{\infty}dt_2 \left\{\left[\bm{q}(t_2)- \bm{q}_i
\right]^2 - \left[\bm{q}_f- \bm{q}_i \right]^2\right\}.
\end{eqnarray}
(Note that in deriving this limit we did not expand the sine in powers of $\omega$, thus avoiding spurious divergences.)

In $I_2$, the leading contribution comes from $\int_{T}^{\infty}dt_2
\int_{0}^{T}dt_1\ldots$, since $\omega \int_{0}^{T}dt_2
\int_{0}^{t_2}dt_1\ldots=\mathcal{O}(T^2/l_f^2)$ (being of a higher
order of smallness), because there, for finite integration limits
and $T/l_f\to0$, the sine in the integrand can be linearized,
bringing an extra $T/l_f$ factor. In the band $0<t_1<T$, $t_2>T$ (a
region symmetrical to that making leading contribution to $I_1$), we
can linearize the phase by omitting term $\frac{t_1t_2}{l_f
\tau}\frac{q_{fi}^2}{m^2}$ (now because for finite $t_1$,
$\frac{t_1t_2}{l_f
\tau}\frac{q_{fi}^2}{m^2}\underset{l_f\to\infty}\to0$), and in the
limit $l_f\to\infty$ get
\begin{eqnarray}\label{Iiet2>T}
I_2\simeq\omega\frac{e^2}{\pi}\frac{E^2+E'^2}{4E^3E'}
\int_{0}^{T}dt_1 \left[\bm{q}_f- \bm{q}(t_1)\right]^2
\int_{T}^{\infty}\frac{dt_2}{\tau}  \sin\frac{\tau}{l_f(\omega)}
\nonumber\\
=\omega\frac{e^2}{\pi} \frac{E^2+E'^2}{4E^3E'} \int_{0}^{T}dt_1 \left[\bm{q}_f- \bm{q}(t_1)\right]^2 \int_{(T-t_1)/l_f(\omega)}^{\infty}\frac{d\varphi}{\varphi}  \sin \varphi \nonumber\\
\underset{T/l_f(\omega)\to0}\to \omega\frac{e^2}{2}
\frac{E^2+E'^2}{4E^3E'} \int_{0}^{T}dt_1 \left[\bm{q}_f-
\bm{q}(t_1)\right]^2\qquad\qquad\quad
\end{eqnarray}
(again, avoiding troublesome expansion of $\sin\varphi$ to power series).

Combining (\ref{Iee-IBH}) and (\ref{Iiet2>T}), on account of
identity
\[
(\bm{q}-\bm{q}_i)^2+(\bm{q}_f-\bm{q})^2-(\bm{q}_f-\bm{q}_i)^2
=-2(\bm{q}_f-\bm{q})\cdot(\bm{q}-\bm{q}_i),
\]
we are led to
\begin{equation}\label{dIdomega-O(omega)}
\frac{dI}{d\omega}\underset{T/l_f(\omega)\to0}\simeq
\frac{dI_{\text{BH}}}{d\omega}\left(\frac{q_{fi}}{m},\frac{\hbar\omega}{E}\right)+C_1\omega+\mathcal{O}\left[T^2/l_f^2(\omega)\right]
\end{equation}
with
\begin{subequations}\label{C1}
\begin{eqnarray}
C_1=-e^2 \frac{E^2+E'^2}{4E^3E'} \int_{-\infty}^{\infty} dt [\bm{q}(t)-\bm{q}_i]\cdot [\bm{q}_f-\bm{q}(t)]\qquad\quad\label{C1-q}\\
=-e^2 \frac{E^2+E'^2}{4E^3E'} \int_{-\infty}^{\infty} dt
\int_{-\infty}^{t} dt'\bm{F}_{\perp}(t')\cdot
\int_{t}^{\infty}dt''\bm{F}_{\perp}(t'').\nonumber\\
\label{C1-F}
\end{eqnarray}
\end{subequations}
Here the lower and upper integration limits were replaced by infinity,
presuming the integrand to vanish rapidly enough at $t<0$ and $t>T$.
Let us stress that expression (\ref{C1}), being quadratic in $q$ (or
$F_{\perp}$), nonetheless implies no restrictions on ratio $q/m$,
i.e., on the dipole or nondipole character of the radiation.
Evidently, integral (\ref{C1}) takes into account the smoothness of
the crossover between the rectilinear asymptotes of the electron
trajectory.

For $\hbar\omega\ll E$, Eq. (\ref{C1-q}) reduces to
\begin{equation}\label{C1-semicl}
C_1=-\frac{e^2}2  \int_{-\infty}^{\infty} dt
[\bm{v}(t)-\bm{v}_i]\cdot [\bm{v}_f-\bm{v}(t)],
\end{equation}
i.e., for a given trajectory it does not depend on the electron
Lorentz-factor; that can be attributed to the radiophysical character of the radiation process in this limit (cf. \cite{Bond-Shul-double-scat}).

To precisely understand the physical meaning of the obtained result,
note that structure (\ref{C1}), (\ref{C1-semicl}) coincides with
that of the second term in the right-hand side (rhs) of
Eq.~(\ref{vtau-chi-chi-b}). Since the meaning of its left-hand side
is clear enough (being a time delay due to the trajectory
curvature), it remains to figure out the meaning of the first term
in the rhs at $t_1\to-\infty$, $t_2\to\infty$. To this end,
integrating by parts as in Eq. (\ref{t1t2+O}), rewrite it as
\begin{eqnarray}
\int_{t_1}^{t_2}dt
\left[\bm{v}_f-\bm{v}(t)\right]\cdot
\int_{t_1}^{t_2}dt \left[\bm{v}(t)-\bm{v}_i\right]\qquad\qquad\qquad\qquad\nonumber\\
\underset{t_2\to\infty}{\underset{t_1\to-\infty}\simeq}-\left(t_1\bm{v}_{fi}-\int_{-\infty}^{\infty}dtt\frac{d\bm{v}}{dt}\right)\cdot
\left(t_2\bm{v}_{fi}-\int_{-\infty}^{\infty}dtt\frac{d\bm{v}}{dt}\right).\nonumber\\
\end{eqnarray}
Choosing the zero time such that
$\int_{-\infty}^{\infty}dtt\frac{d\bm{v}}{dt}=0$ (which is always
possible, e.g., when the motion is planar), integrals
$\int_{t_1}^{t_2}dt \left[\bm{v}_f-\bm{v}(t)\right]$ and
$\int_{t_1}^{t_2}dt \left[\bm{v}(t)-\bm{v}_i\right]$ (representing
the particle transverse coordinates with respect to $\bm{v}_f$ or $\bm{v}_i$) at $t_1\to-\infty$, $t_2\to\infty$ are proportional correspondingly to $t_1$ and $t_2$, i.e., both the initial and the final
trajectory asymptotes issue from the origin. Correspondingly, in
that limit the first term in the rhs of Eq.~(\ref{vtau-chi-chi-b})
tends to the time delay $v(t_2-t_1)-|\bm{r}(t_2)-\bm{r}(t_1)|$ for a
trajectory having the shape of an angle along the initial and final
electron asymptotes. The second term thus represents a difference
between the time delay for the actual trajectory and for its
angle-shaped approximation. If it is impossible to adjust the time
origin such that $\int_{-\infty}^{\infty}dtt\frac{d\bm{v}}{dt}=0$,
it suffices to demand that
$\bm{v}_{fi}\cdot\int_{-\infty}^{\infty}dtt\frac{d\bm{v}}{dt}=0$.

Generally, for monotonous electron deflection, $C_1\leq 0$ (the
particle ``cuts the corner"), whereas for an oscillatory electron
motion within the target, $C_1\geq 0$. As a cross-check, note that
for classical radiation at double scattering through angles
$\bm{\chi}_1$ and $\bm{\chi}_2$ with a time separation $t_{21}$,
Eq.~(\ref{C1-semicl}) gives
\[
C_1=-\frac{e^2}{2}\bm{\chi}_1\cdot\bm{\chi}_2 t_{21},
\]
which coincides with the result obtained in
\cite{Bond-Shul-double-scat}. After an initial decline, however, the spectrum will start rising, due to resolution of smaller parts of the electron
trajectory.

In case of a monotonous electron deflection, as in a magnet of length $T$, $C_1$ grows with $T$ cubically, wherefore at low $\omega$ it may compete with the ``volume" (synchrotron-like) contribution, which is proportional to $T$.

In case of undulator radiation, when $\bm{F}_{\perp}(t)=\bm{F}_0\cos\frac{2\pi t}{T_1}$
within the interval $0<t<NT_1$, with $N\gg1$ being the number of oscillation periods, from (\ref{C1-F}) we get
\begin{equation}\label{C1/NT1}
\frac{C_1}{NT_1}\underset{N\to\infty}\simeq
e^2\frac{E^2+E'^2}{8E^3E'}\left(\frac{F_0 T_1}{2\pi}\right)^2.
\end{equation}

Finally, in an amorphous medium modeled by action of a delta-correlated (Langevin) force, averaging of (\ref{C1-F}) gives
zero:
\begin{eqnarray}
\left\langle \int_{-\infty}^t dt'\bm{F}_{\perp}(t')\cdot \int_t^{\infty} dt''\bm{F}_{\perp}(t'') \right\rangle \nonumber\\
\propto \int_{-\infty}^t dt'  \int_t^{\infty} dt''\delta(t'-t'') =0.
\end{eqnarray}

\section{Summary}

The distinctions of our result (\ref{C1}) from the Low theorem
\cite{Low} are that it applies to: (i) the angle-integral radiation
spectrum, which is a more inclusive quantity than the radiation
amplitude considered in \cite{Low}; (ii) ultrarelativistic electrons
and small-angle photon emission, allowing for $\hbar\omega\sim E$,
whereas the small parameter is $T/l_f(\omega)$. From the physical
point of view, the key notion is that the NLO expansion for
$dI/d\omega$ begins with $\mathcal{O}(T/l_f)$, since in the double
time integral representation (\ref{dIdomega-t1t2}) it originates
from the region where only one of the contributing times is large:
$|t_1|$ for $I_1$ and $t_2$ for $I_2$.


The obtained correction, by virtue of its simplicity, can be used
for accurate connection of the infrared (or generalized
factorization) limit of the bremsstrahlung spectrum with its
behavior at higher $\omega$, probing the interior of the target. At
its application, it is worth minding that the correction is
insensitive to nondipole radiation effects. For instance, relation
(\ref{C1/NT1}), well known for dipole undulators, must hold as well
for wigglers.

\subsection*{Acknowledgements}

This work was supported in part by the Ministry of Education and
Science of Ukraine (Project No. 0115U000473).

\end{document}